\input epsf.tex
\input psfig.sty
\newcommand{\sperp}{{\scriptscriptstyle\perp}}
\newcommand{\spm}{{\scriptscriptstyle\pm}}

\documentstyle[prl,aps]{revtex}
\begin{document}
\twocolumn[\hsize\textwidth\columnwidth\hsize\csname@twocolumnfalse%
\endcsname

\draft
\rightline{McGill 97-37}
\title{Electromagnetic Response and Approximate SO(5) 
Symmetry \\ in High-$T_c$ Superconductors}

\author{C.P.~Burgess, J.M.~Cline}

\address{Physics Department, McGill University,
3600 University Street, Montr\'eal, Qu\'ebec, Canada H3A 2T8.}

\author{C.A.~L\"utken}
\address{Physics Department, University of Oslo,
P.O. Box 1048, Blindern, N-0316 Norway.}

\maketitle

\begin{abstract}
\end{abstract}
{It has been proposed that the effective Hamiltonian describing high
$T_c$ superconductivity in cuprate materials has an approximate $SO(5)$
symmetry relating the superconducting (SC) and antiferromagnetic (AF)
phases of these systems.  We show that robust consequences of this
proposal are potentially large optical conductivities and Raman 
scattering rates in the AF phase, due to the electromagnetic response of the
doubly-charged pseudo Goldstone bosons which must exist there.  
This provides strong constraints on the properties of the bosons, such as
their mass gap and velocity.
}
]

%
%

It has recently been proposed \cite{zhang-sci} that the
antiferromagnetic (AF) and superconducting (SC) phases of the
high-$T_c$ cuprates are related by an $SO(5)$ symmetry, a suggestion
which has both stimulated considerable interest
\cite{others,vortices,sandwich,pGBproperties} and drawn sharp criticism
\cite{criticism,anderson}.  Both the interest and the criticism are
inspired by the central role played by the symmetry relating the two
phases.

The $SO(5)$ picture has much to recommend it.  It gives a simple and
concrete qualitative understanding of many features of the cuprates, as
well as predicting brand new phenomena, such as superconducting
vortices having antiferromagnetic cores \cite{vortices} and persistent
superconducting phase correlations within the AF phase \cite{sandwich}.
On the negative side, Anderson and Baskaran have argued that such a
finite-dimensional symmetry description of the cuprates is inconsistent
with the electron localization properties of the two phases
\cite{anderson}.

In this work we exploit another strength of the $SO(5)$ theory --
namely its predictive power -- to bring further evidence to the
discussion. A model-independent consequence of the $SO(5)$ picture is
the existence in the AF phase of an electrically charged
pseudo-Goldstone boson (pGB) quasiparticle, whose dispersion relation
and low-energy couplings are tightly constrained by the $SO(5)$
symmetry \cite{zhang-sci, pGBproperties}. We use these to compute the
contribution of these quasiparticles to the electromagnetic response of
the cuprate materials in the AF phase. In particular we find the 
Raman scattering rate and the real part of the far-infrared conductivity,
which turn out to be comparable to or larger than what is
experimentally observed.  We emphasize that our calculation relies
almost exclusively on the assumed $SO(5)$ symmetry, and depends only
minimally on the microscopic details of these systems.

The key tool in the analysis is the effective Lagrangian density
which describes the electromagnetic couplings of the $SO(5)$
pGB's. This can be written as 
\begin{eqnarray}
{\cal L}&=& \left|\left({\partial_t}-2ie(\mu + A_0)\right)\phi\right|^2  - 
  \sum_i \left|v^\phi_i \left({\nabla_i} -{\textstyle{ 2ie\over c}}
	A_i\right)
  \phi\right|^2  \nonumber\\
\label{hamiltonian}
   \qquad && +  \left( {\partial_t {\vec n}} \right)^2
 -  \sum_i \left(v^n_i {\nabla_i} \vec n\right)^2 - V(\phi,\vec{n}).
\end{eqnarray}
Here $\mu$ is the chemical potential which describes the system's
doping, $A_i$ is the electromagnetic gauge potential, and 
$v_i^\phi$ and $v_i^n$ are the pGB velocities,
which can differ along the three principal directions of the medium.
In the limit of exact SO(5) symmetry, we would have $v^n_i=
v^\phi_i \equiv v_i$, and the scalar potential $V(\phi,\vec{n})$ 
satisfying $V = V(|\phi|^2 + {\vec{n}}^2)$. 

There are two important, but conceptually very different, regimes to which
this lagrangian applies. 

1. Deep within the AF or SC phases, where fluctuations in the modulus
$|\phi|^2 + {\vec{n}}^2$ are negligible, $V$ becomes a constant in the
SO(5) symmetry limit, and $\phi$ and $\vec n$ describe the dynamics of
the Goldstone and pseudo-Goldstone quasiparticles.  For instance for
the AF phase there are four such modes: two gapless magnons with
dispersion relation $E^2_n(p) = \sum_i (v^n_i \, p_i)^2$, and two pGB's
with charge $\pm 2e$ and dispersion $(E_\phi(p) \mp 2 e\mu)^2 =
\varepsilon_g^2 + \sum_i (v^\phi_i \, p_i)^2$. Approximate $SO(5)$
invariance amounts to the statement that $|v^\phi_i - v^n_i|$ is small
compared to $v_i$, and the gap, $\varepsilon_g$, is much smaller than
the typical microscopic scale of the system: $J \sim 0.1$ eV.  `Much
smaller' here means of comparable size to the experimentally-measured
41 meV gap \cite{41mev} which is interpreted within the $SO(5)$ context
as a pGB of the SC phase.  Here the two-dimensional nature of the
cuprates dictates the sum on $i$ runs only over the two spatial
derivatives ($x$ and $y$) which label the copper-oxygen planes.

2. Near the critical boundaries between the various phases, $\cal L$
plays the role of a Ginzburg-Landau (GL) free energy, and (in mean
field theory) $V$ may be expanded to quartic order:  $V = -
m^2_\phi|\phi|^2 - {\scriptstyle\frac12} m^2_n|\vec n|^2
+ \lambda_\phi|\phi|^4 + 2\lambda_{\phi n}|\phi|^2|\vec n|^2 + 
\lambda_n|\vec n|^2$.  In this
case the model can be two- or three-dimensional (but anisotropic, $v_z
\ne v_x = v_y$) depending on how close one is to the critical limit.
$SO(5)$ invariance implies in this case $m^2_\phi = m^2_n$ and
$\lambda_\phi= \lambda_n=\lambda_{n\phi}$, and so approximate $SO(5)$
invariance is the statement that the deviations from these relations
are systematically small. Phenomenology requires $m^2_n > m^2_\phi$
when the doping is sufficiently small, in order that the ground state
of the system be AF, with $\phi=0$ and $\vec n\neq 0$. In this regime
the pseudo-Goldstone gap in the AF phase may be computed, giving
$\varepsilon_g^2 = -m^2_\phi + {\lambda_{\phi n}\over\lambda_n}m^2_n$,
which clearly vanishes in the $SO(5)$ limit, as required.

The existence of charged bosons with a small gap has strong observable
consequences for the optical conductivity and Raman scattering
properties in the AF phase.  These predictions are quite robust.  The
interaction of the photons with the bosons, the first two terms of
eq.~(\ref{hamiltonian}), is completely fixed by electromagnetic gauge
invariance. The self-interactions described by the other terms,
including possible higher powers of $|\phi|$ not shown, must vanish for
Goldstone bosons in the limit of zero energy and exact $SO(5)$ symmetry,
and so may be treated perturbatively for low energies and
approximate symmetry. These also ensure the small size of the gap,
$\varepsilon_g$.

{\it Conductivity:} We start with the contribution of the charged
pseudo-Goldstone quasiparticle to the real part of the conductivity, 
$\sigma_1$, within the AF phase.  This is proportional to the imaginary part
of the photon self-energy due to a virtual pseudo-Goldstone boson loop,
and related to the rate at which the pGB's are pair-produced by
incident photons.  Our result agrees in the
appropriate limits with closely analogous formulas for the
electromagnetic response of a hot pion gas \cite{pions}.  For pGB's
which can move in three dimensions, and whose velocity is much less 
than that of light, we obtain
\begin{equation}
\label{cond3d}
\sigma_{\rm1,i} = 
{e^2\over 24\pi\hbar\lambda} 
\left({v_i^2 c\over v_x v_y v_z}\right)\!
\left( 1 - {4\varepsilon_g^2 \over \hbar^2\omega^2} 
\right)^{3/2}\!\! \Bigl[ 1 + N_{\scriptscriptstyle B} \left( {\omega
\over 2} \right) \Bigr],
\end{equation}
where the absorbed light of wavelength $\lambda$ is polarized in the
`$i$' direction, taken to be one of the crystal axes.  (Henceforth we
drop the superscript $\phi$ from the pGB velocities $v_i$.)
$N_{\scriptscriptstyle B}(\omega) = n_+(\omega) + n_-(\omega)$, where
$n_\pm(\omega) = 1/(\exp[ (\hbar\omega \pm \mu)/kT ] - 1)$ is the usual
Bose-Einstein statistics factor, and $e^2/\hbar = 2.4341\times
10^{-4}\ \Omega^{-1}$.  This contribution to the conductivity vanishes
for photons whose energy $\hbar\omega$ is below the threshold
$2\varepsilon_g$ for producing two pGB's.  The corresponding absorption
coefficient is given by $\sigma_{\rm1,i}/(\epsilon_0 c)$, where 
$\epsilon_0 c = 2.654\times 10^{-3} \Omega^{-1}$.


%

To take into account that the pGB's may be confined to move in the
superconducting planes, we have also computed $\sigma_1$ using the
lattice dispersion relation $E = (v_\sperp^2 p_\sperp^2 +
(2 v_z\sin(ap_z/2)/a)^2)+\varepsilon_g^2)^{1/2}$, 
integrating $p_z$ between
$\pm\pi/a$, where $a$ is the spacing between the planes.  Defining
$\eta = 1-\frac12(a/v)^2(\omega^2 - \varepsilon_g^2/\hbar^2)$, 
the phase space integral can be done exactly, giving 
\begin{equation}
\label{cond2d}
\sigma_{1,i} = {e^2\over 16\pi^2\hbar a}\left({v_i v_z\over 
v_\sperp a\omega} \right)^2 
 \Bigl[ 1 + N_{\scriptscriptstyle B} \left( {\omega
\over 2} \right) \Bigr] G_i(\eta),
\end{equation}
where $G_\sperp(\eta) = \sqrt{1-\eta^2} - \eta \cos^{-1}\eta$ 
 and
$G_z(\eta) = \frac12(\cos^{-1}\eta - \eta\sqrt{1-\eta^2} )$ (
these become respectively $-\pi\eta$ and $\pi/2$
for $\eta<-1$).  The continuum
approximation on which the 3D equation (\ref{cond3d}) is based is valid
when $a\to 0$ (so $\eta\to 1$).  For the parameters of interest, we find
that the 3D formula is practically indistinguishable from eq.~(\ref{cond2d}),
although the difference starts to become apparent for frequencies greater than
800 cm$^{-1}$.

\vspace{0.25in}
\centerline{\epsfxsize=3in\epsfbox{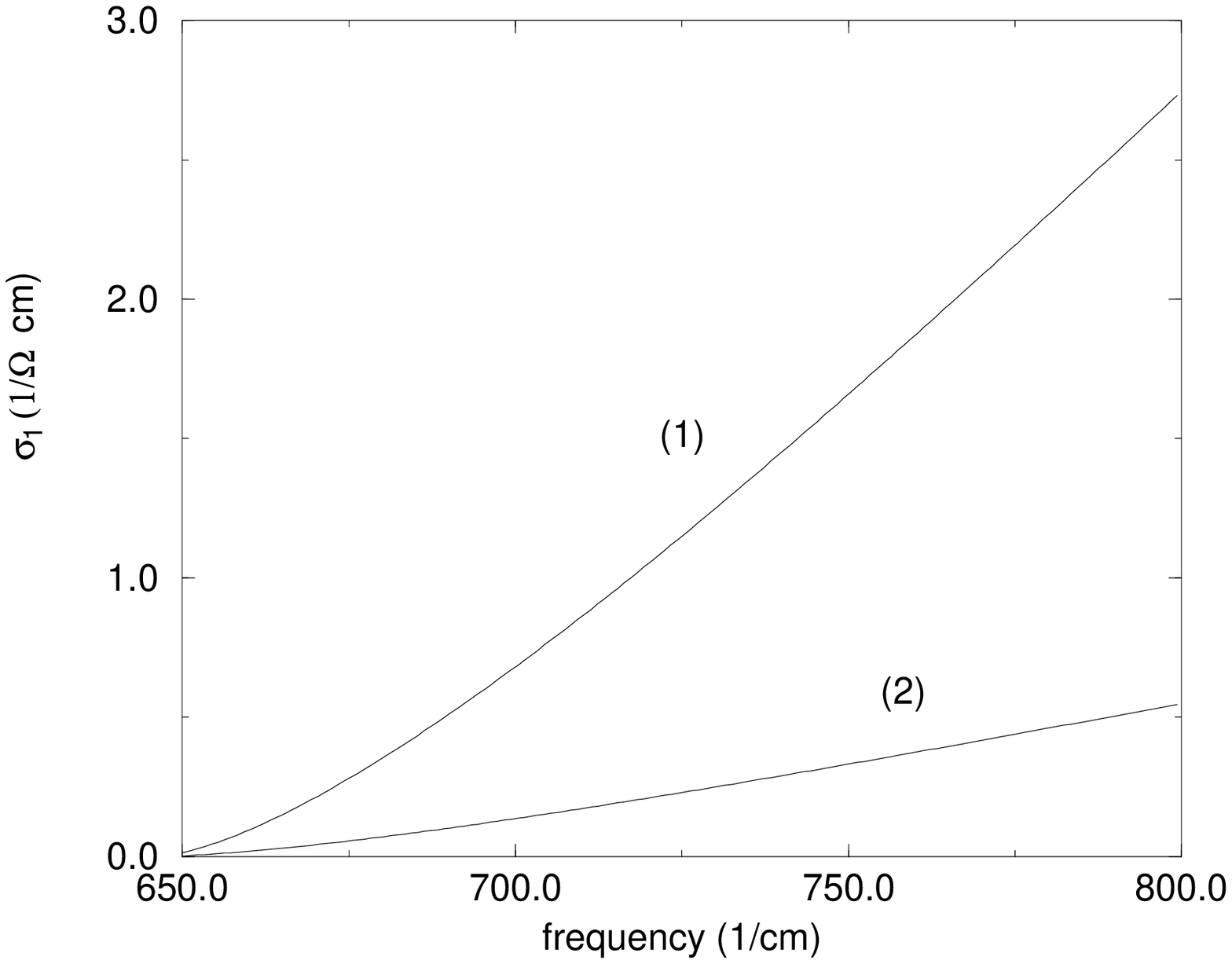}}
\medskip\noindent
{\small Figure 1:
The optical conductivity ${\sigma_1}$ as a function of the photon energy
$\omega$, in units of $\Omega^{-1}$cm$^{-1}$, for two different pGB
velocities, (1) $v/c = 2\times 10^{-4}$, and (2) $v/c = 10^{-3}$.
The other relevant parameters
are $\varepsilon_g = 0.04$ eV, $T = 300$ K, and $\mu = 0$.}
\medskip

How big are these results?  If the gap is approximately
$\varepsilon_g=0.04$ eV as suggested by the neutron scattering data,
the interplane spacing is $a=0.5$ nm, and using the magnon
speed $v_i = 2\times 10^{-4}c$ \cite{magnonspeed},
then for $\omega= 0.1$ eV (frequency $=800$ cm$^{-1}$) and $T = 0.025$
eV (300K), $\sigma_1 = 2.75 \Omega^{-1}$cm$^{-1}$.  This is more
than three times larger than the measured values of $\sigma_1 <
1\ \Omega^{-1}$cm$^{-1}$ for the undoped cuprate Sr$_2$CuO$_2$Cl$_2$,
as reported in reference \cite{Tanner}; moreover the frequency
dependence disagrees with the data, which has $\sigma_1$ decreasing in
this frequency range.  Although the perturbative treatment of the pGB's
begins to break down at higher energies, if we can trust our results at
somewhat higher energies of $\omega = 0.5$ eV ($=4000$ cm$^{-1}$), then
$\sigma_1$ grows to 100 $\Omega^{-1}$cm$^{-1}$, in even greater
conflict with measured values near 1 $\Omega^{-1}$cm$^{-1}$ in the
materials Gd$_2$CuO$_4$ and YBa$_2$Cu$_3$O$_6$ \cite{Tanner}.
(Reference \cite{Timusk} has also measured conductivities, but their
maximum frequency of 700 cm$^{-1}$ may be below the threshold
$2\varepsilon_g$ needed for production of pGB pairs.)

Although the magnitude and the shape of $\sigma_1$ suggest a conflict
between SO(5) and experiments, there are several caveats.  One is that
$\sigma_1$ is inversely proportional to the pGB velocity, $v_i$, which is not
precisely known.  The other is that our computation is only valid for
energies within the domain of approximation of the low-energy pGB
lagrangian used here, {\it i.e.} for $\omega \ll J \sim 0.1 $eV. Above
these energies the pGB's need not contribute as a weakly-coupled and
comparatively narrow state.  The experimental constraints happen to
be strongest just in the region where our long-wavelength approximation
starts to break down.  Thus we turn to another possible signal of
the pGB electric response.

{\it Raman Scattering:}
We now consider the contribution to the Raman scattering rate, 
coming from Compton-like photon scattering from the pseudo-Goldstone
quasiparticles in the sample. The Feynman graphs for this
process are those of Figure 2.

\vspace{0.25in}
\centerline{\epsfxsize=3in\epsfbox{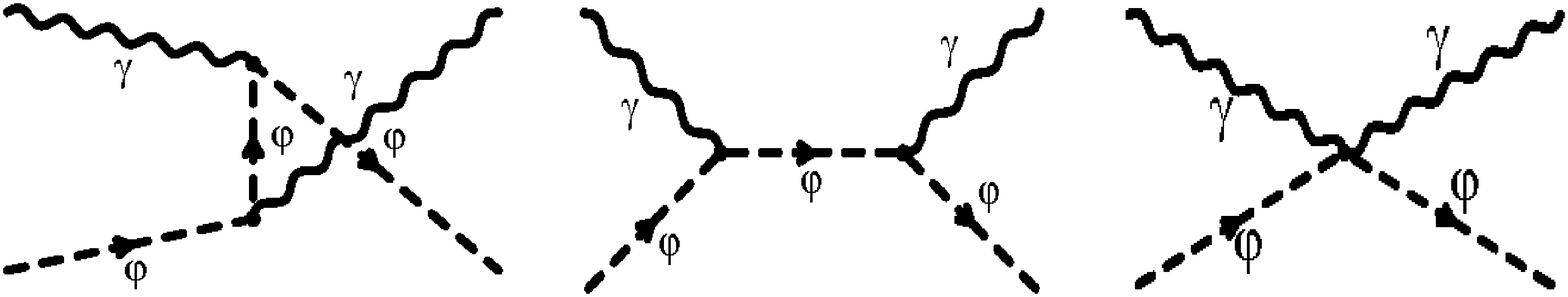}}
\centerline{ (a) $\qquad\qquad\qquad$ (b) $\qquad\qquad\qquad$ (c) }
\medskip\noindent
{\small Figure 2:
The Feynman graphs which describe photon-pGB scattering.}
\medskip

We work in a gauge for which the expression
for the scattering amplitude, ${\cal M}$, is particularly simple. That
is, if $p_{i,f}^\mu$ and $k_{i,f}^\mu$ denote the initial and final pGB
and photon four-momenta, respectively, and we choose the photon
polarizations, $\tilde\epsilon_{i,f}^\mu$, to satisfy $\tilde
\epsilon_i \cdot (2 \tilde p_i + \tilde k_i) = \tilde\epsilon_f \cdot
(2 \tilde p_i - \tilde k_f) = 0$, then diagrams (2a) and (2b) of the
Figure vanish, leaving an amplitude of the form ${\cal M}_{ab} =
4e^2\tilde\varepsilon^{(a)}_i\!\cdot \tilde\varepsilon^{(b)}_{f}$.  We
use here the convenient notation for any four-vector, in which a
`$\tilde{\phantom{\varepsilon}}$' indicates the multiplication of the
spatial components by the corresponding velocity, $v_i$. That is,
$\tilde p_0 = p_0$, and $\tilde p_i = v_i p_i$.

Let us assume that the initial photon is moving in the $z$ direction,
perpendicular to the superconducting planes, and
the final one is scattered by 180$^\circ$, as is the case in 
the experiments to which we compare.  Assuming the $x$-$y$
plane to be isotropic,  $v_x=v_y=v_\sperp$, and 
averaging over initial and summing over final polarizations 
leads to the following expression:
\begin{equation}
{\cal S} \equiv  \frac12\sum_{a,b}|{\cal M}_{ab}|^2 = 8e^4 v_\sperp^4 
\Bigl[ 1 +(1 - A)^2  \Bigr], 
\end{equation}
where $A = (1+v_z^2) v_\sperp^2 p_{\sperp}^2/(D_i D_f)$, with
\begin{eqnarray}
   D_i &=& E_i - v_z^2 p_{i,z}+ \frac12 \, \omega_i(1 - v_z^2) , \\
 D_f &=& E_i + v_z^2 p_{i,z} - \frac12 \, \omega_f(1 - v_z^2) .
\end{eqnarray}
Conservation of four-momentum constrains the initial and final 
boson energies and momenta according to
\begin{eqnarray}
 \label{init_cond}
 E_i &=&\phantom{-}\frac{\Delta\omega}{2}+\omega_0 v_z f(p_{\sperp});
\quad
 p_{i,z} = -\omega_0 - \frac{\Delta\omega}{2v_z}f(p_{\sperp})\nonumber\\
 \label{final_cond}
 E_f &=& -\frac{\Delta\omega}{2} + \omega_0 v_z f(p_{\sperp});\quad
 p_{f,z} = \phantom{-}\omega_0 - \frac{\Delta\omega}{2v_z}f(p_{\sperp}),
\end{eqnarray}
where $\Delta\omega = \omega_f-\omega_i$, 
$\omega_0 =(\omega_i+\omega_f)/2$,
and 
\begin{equation}
f(p_\sperp) = \left(1+{\varepsilon_g^2+v_\sperp^2 p_\sperp^2 \over 
|v_z^2\omega_0^2 - 
\Delta\omega^2/4|}\right)^{1/2}.
\end{equation}
Care must be taken with these expressions, however, because the
quantity $A$ can diverge for some value of $p_{\sperp}$.  In a general
gauge this comes about when the virtual boson in diagram (2b) goes on
its mass shell.  (Equivalently, the gauge transformation which we used
to remove diagram (2b) becomes singular for these momenta.) We handle
this situation, when it arises, by including the width of the pGB
itself. This may be done by adding a small imaginary part, $i\Gamma/2$,
to the boson energies in the definition of $A$, giving a finite
scattering rate. We find our results to be insensitive to reasonable
values such that $\Gamma\le \varepsilon_g$, for relevant temperatures
and energies.

The observable of interest is the differential scattering rate per
unit sample thickness, $l$, per unit incident laser power, $I$: $R =
(1/I) (d\Gamma / dl)$.  This is the quantity which does not depend on
the details of the target or of the incident photon flux.  We compute
the differential rate of such scattering into a solid angle $d\Omega$
(centred about 180${}^\circ$), and into a final energy interval
$d\omega_f$, ${\cal R} = dR/d\omega_f d\Omega$. We then do the
thermal average over the initial and final pGB's, sum
over the final photon polarizations and average over the initial ones. 
This assumption that the incident photons are unpolarized, and
scattered-photon polarizations are not detected, simplifies our
expressions, but is not crucial for the result.  We find
%
\[
   {\cal R}\equiv {\omega_f\over (4\pi)^4 v_\sperp^2 \omega_i^2
	|\Delta\omega|}
	\sum_\spm\int { d p_{z} }
   n_\pm(E_i)(1+n_\pm(E_f)){ \cal S}
\]
%
where $n_\spm$ is the Bose-Einstein distribution 
function with chemical
potential $\pm\mu$.  The limits of
integration are found by varying $p_\sperp^2$ between 0 and $\infty$ in
eq.~(\ref{init_cond}) for $p_{i,z}$, and $p_\sperp^2$ in the integrand
is determined by $p_{i,z}$ by inverting the same equation.
As a function of
the final photon energy, the rate is peaked near $\omega_f=\omega_i$,
with an approximate maximum value of 
\begin{equation}
\label{Rmax}
{\cal R}_{\rm max} \cong \left({\alpha v_\sperp\over\pi v_z\omega}
\right)^2 T e^{-\beta(\varepsilon_g-|\mu|)},
\end{equation}
using the Boltzmann approximation for the distribution functions.
Kinematics constrains the scattered photon energy to lie in the narrow
range ${1-v_z\over 1+v_z} < {\omega_f\over\omega_i} < {1+v_z\over
1-v_z}$.  For visible light (500 nm) and $v=10^{-3}c$, this gives a
half-width of 40 cm$^{-1}$ in the frequency $\nu$.  In figure 3 we show
${\cal R}$ for representative input values 
$\omega_i = 2$ eV, $T = 93$ K, $\varepsilon_g = 0.04$ eV, $\mu = 0$ and
$0.01$ eV, $\Gamma= 0.1\varepsilon_g$, and optimistically large velocities
$v_z=v_{\sperp}=10^{-3}c$.  For these numbers we find rates of order 
${\cal R}\sim (150-300)$
cps/mW/\AA/steradian/eV, with a smooth, featureless lineshape.  For
optical photons the penetration depth of typical samples is of order
100 \AA \cite{Hackl}, leading to a prediction of $10^4$ 
events/mW/\AA/sr/eV in the backward direction.  
Experimentally, the observed rate for the compound Bi2212 is
only 15 cps/mW/\AA/sr/eV.  However this must be corrected 
\cite{Hackl_private} for
detector efficiency (0.1) and surface losses ($0.1- 0.01$), which
in the worst case would bring the predicted value into agreement
with the experiments.

We point out that the predicted rate is exponentially dependent on the
ratio of the pGB gap or chemical potential to the temperature because
of the Boltzmann factor in eq.~(\ref{Rmax}).  If
$(\varepsilon_g-|\mu|)\le kT$, there is a further gain by a factor of
150 in the rate, which would cause a serious discrepancy between the
predicted and observed values.  On the other hand, the width depends
linearly on the pGB velocity.  If $v < 2\times 10^{-4}c$, this width
becomes less than the experimental resolution of ref.~\cite{Hackl},
which only measures Raman shifts greater than $\sim 10$/cm.

\centerline{\epsfxsize=8.5cm\epsfbox{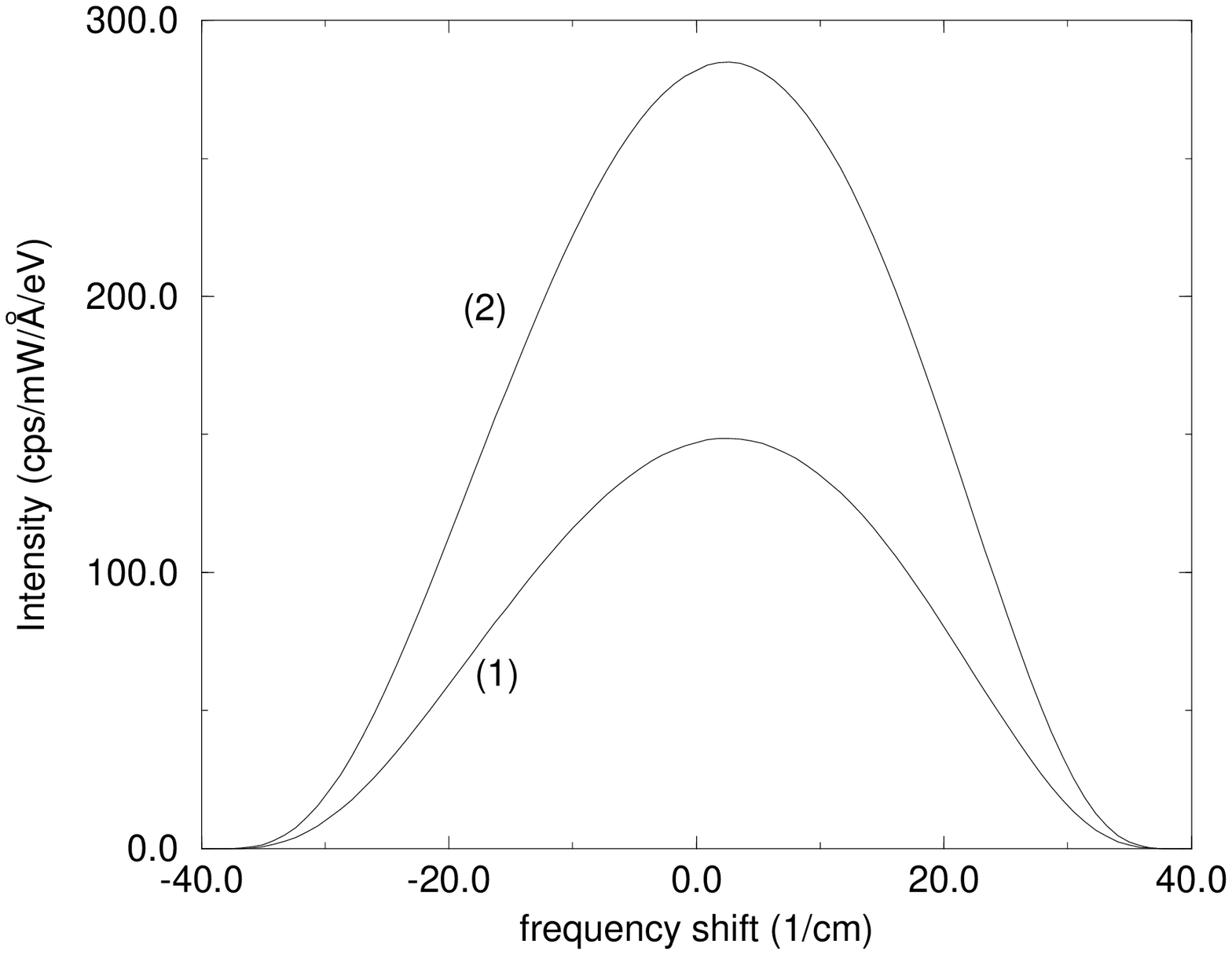}}
\vspace{-0.25in}
\medskip\noindent
{\small Figure 3:
The differential Raman scattering rate ${\cal R}$ as a function of the
photon frequency shift $\nu_i-\nu_f$, in units of cm$^{-1}$.  The
curves correspond to: (1) the parameters given in the text; and 
(2) same as (1) except with nonzero chemical potential,
$\mu = 0.01$ eV.} 
\medskip

Similarly to the conductivity, the Raman scattering rate near zero
shift depends only weakly on whether the pGB's are allowed to move
in three dimensions or confined to the 2D planes, for the parameters of
interest.  By repeating the above calculation using the lattice
dispersion relation for the pGB's, one finds that the continuum version
is a good approximation when $\omega a\ll c$, rather than the condition
$\omega a \ll v$ which applied for the conductivity.  Although we are 
interested in larger frequencies in Raman scattering than for the
conductivity, the latter condition is still satisfied, and the differences
between the 2D and 3D Raman intensities are small.

In conclusion, we have used $SO(5)$ symmetry to compute the low-energy
contribution of the electrically-charged pseudo-Goldstone bosons to the
electromagnetic response of cuprates doped to be antiferromagnets. We
find measurably large conductivities and Raman scattering rates, due to
the presence of electrically-charged states with a comparatively small
gap.  At present the dispersion relation of the putative bosons is not
sufficiently well known to rule out SO(5) for the cuprates based on the
data.  It is encouraging, however that the conductivity and Raman
intensities have a complementary dependence on the pGB velocity, so
that one or the other should show evidence for the pGB's, especially if
the experiments are improved, {\it e.g.} Raman scattering at frequency
shifts less than 10/cm.

We thank R.~Hackl and T. Timusk for information about the 
experiments, and C.~Gale and A.~Berlinksy for useful discussions.

\vskip-0.5cm

\end{document}